\documentclass[twocolumn,english]{IEEEtran}
\usepackage[T1]{fontenc}
\usepackage[latin9]{inputenc}
\usepackage{geometry}
\usepackage{color}
\usepackage{amsmath}
\usepackage{amssymb}

\makeatletter
\usepackage[caption=false,font=footnotesize]{subfig}

\newtheorem{assumption}{Assumption}
\newtheorem{theorem}{Theorem}

\newtheorem{definition}{Definition}

\makeatother

\usepackage{babel}
\begin{document}

\title{Data-Driven Inversion-Based Control: \\closed-loop stability analysis
for MIMO systems}

\author{Carlo Novara and Simone Formentin\thanks{C. Novara is with Politecnico di Torino (Italy), S. Formentin is with
Politecnico di Milano (Italy). Email: \texttt{\footnotesize{}carlo.novara@polito.it},
\texttt{\footnotesize{}simone.formentin@polimi.it}.}}
\maketitle
\begin{abstract}
Data-Driven Inversion-Based Control (D$^{2}$-IBC) is a recently introduced
control design method for uncertain nonlinear systems, relying on
a two degree-of-freedom architecture, with a nonlinear controller
and a linear controller running in parallel. In this paper, extending
to the MIMO case a previous result holding for the SISO case, we derive
a finite-gain stability sufficient condition for a closed-loop system
formed by a nonlinear MIMO plant, connected in feedback with a D$^{2}$-IBC
controller.
\end{abstract}

\section{Introduction}

Data-Driven Inversion-Based Control (D$^{2}$-IBC) is a recently introduced
control design method for uncertain nonlinear systems, relying on
a two degree-of-freedom architecture, with a nonlinear controller
and a linear controller running in parallel, \cite{formentin2015active,novara2016data}.
Despite many different approaches for joint design of identification
and control have been already proposed (see \cite{hjalmarsson2005experiment}
and the reference therein for a comprehensive overview), as far as
we are aware, D$^{2}$-IBC is the first ``identification for control\textquotedbl{}
method for nonlinear dynamical systems, where also stability guarantees
are provided and enforced directly in the identification algorithm.
However, the underlying stability sufficient condition is valid for
Single Input Single Output (SISO) systems only. In this paper, we
extend this sufficient condition to Multiple Input Multiple Output
(MIMO) systems. \medskip

\textbf{Notation.} A column vector $x\in\mathbb{R}^{n_{x}\times1}$
is denoted as $x=\left(x_{1},\ldots,x_{n_{x}}\right)$. A row vector
$x\in\mathbb{R}^{1\times n_{x}}$ is denoted as $x=\left[x_{1},\ldots,x_{n_{x}}\right]=\left(x_{1},\ldots,x_{n_{x}}\right)^{\top}$,
where $\top$ indicates the transpose.\\
 A discrete-time signal (i.e. a sequence of vectors) is denoted with
the bold style: $\boldsymbol{x}=(x_{1},x_{2},\ldots)$, where $x_{t}\in\mathbb{R}^{n_{x}\times1}$
and $t=1,2,\ldots$ indicates the discrete time; $x_{i,t}$ is the
$i$th component of the signal $\boldsymbol{x}$ at time $t$.\\
 A regressor, i.e. a vector that, at time $t$, contains $n$ present
and past values of a variable, is indicated with the bold style and
the time index: $\boldsymbol{x}_{t}=\left(x_{t},\ldots,x_{t-n+1}\right)$.\\
 The $\ell_{p}$ norms of a vector $x=\left(x_{1},\ldots,x_{n_{x}}\right)$
are defined as 
\[
\left\Vert x\right\Vert _{p}\doteq\begin{cases}
\left(\sum_{i=1}^{n_{x}}\left|x_{i}\right|^{p}\right)^{\frac{1}{p}}, & p<\infty,\\
\max_{i}\left|x_{i}\right|, & p=\infty.
\end{cases}
\]
The $\ell_{p}$ norms of a signal $\boldsymbol{x}=(x_{1},x_{2},\ldots)$
are defined as 
\[
\left\Vert \boldsymbol{x}\right\Vert _{p}\doteq\begin{cases}
\left(\sum_{t=1}^{\infty}\sum_{i=1}^{n_{x}}\left|x_{i,t}\right|^{p}\right)^{\frac{1}{p}}, & p<\infty,\\
\max_{i,t}\left|x_{i,t}\right|, & p=\infty,
\end{cases}
\]
where $x_{i,t}$ is the $i$th component of the signal $\boldsymbol{x}$
at time $t$. 

\section{The D$^{2}$-IBC approach}

\label{sec:ibc_approach}

Consider a nonlinear discrete-time MIMO system in regression form:
\begin{equation}
\begin{array}[t]{l}
y_{t+1}=g\left(\boldsymbol{y}_{t},\boldsymbol{u}_{t},\boldsymbol{\xi}_{t}\right)\vspace{1.5mm}\\
\boldsymbol{y}_{t}=\left(y_{t},\ldots,y_{t-n+1}\right)\\
\boldsymbol{u}_{t}=\left(u_{t},\ldots,u_{t-n+1}\right)\\
\boldsymbol{\xi}_{t}=\left(\xi_{t},\ldots,\xi_{t-n+1}\right)
\end{array}\label{ss_sys}
\end{equation}
where $u_{t}\in U=[-\bar{u},\bar{u}]^{n_{u}}\subset\mathbb{R}^{n_{u}}$
is the saturated input, $y_{t}\in\mathbb{R}^{n_{y}}$ is the output,
$\xi_{t}\in\Xi\doteq\left[-\bar{\xi},\bar{\xi}\right]^{n_{\xi}}\subset\mathbb{R}^{n_{\xi}}$
is a disturbance, and $n$ is the system order. Both $U$ and $\Xi$
are compact sets.

Suppose that the system (\ref{ss_sys}) is unknown, but a set of measurements
is available: 
\begin{equation}
\mathcal{D}\doteq\left\{ \tilde{u}_{t},\tilde{y}_{t}\right\} _{t=1-L}^{0}\label{eq:data}
\end{equation}
where $\tilde{u}_{t}\in U$, $\tilde{y}_{t}\in Y$, $Y=[-\bar{y},\bar{y}]^{n_{y}}$,
$\bar{y}\doteq\max_{t,i}|\tilde{y}_{t,i}|<\infty$, and $\tilde{y}_{t,i}$
is the $i$th component of $\tilde{y}_{t}$. The tilde is used to
indicate the input and output samples of the data set (\ref{eq:data}),
which is supposed to be available at time $t=0$ when the controller
needs to be designed. The input signals employed to generate (\ref{eq:data})
are also assumed to be such that the system output does not diverge.

Let $R\doteq[-\bar{r},\bar{r}]^{n_{y}}$, with $0\leq\bar{r}\leq\bar{y}$,
be a domain of interest for the trajectories of the system (\ref{ss_sys}).
The aim is to control the system (\ref{ss_sys}) in such a way that,
starting from any initial condition $\boldsymbol{y}_{0}\in\mathcal{Y}^{0}\doteq R^{n\times n_{y}}\subset\mathbb{R}^{n\times n_{y}}$,
the system output sequence $\boldsymbol{y}=(y_{1},y_{2},\ldots)$
tracks any reference sequence $\boldsymbol{r}=(r_{1},r_{2},\ldots)\in\mathcal{R}\subseteq R^{\infty}\subset\ell_{\infty}$.
The set of all possible disturbance sequences is defined as $\varXi\doteq\left\{ \boldsymbol{\xi}=(\xi_{1},\xi_{2},\ldots):\xi_{t}\in\Xi,\forall t\right\} $.

To accomplish this task, we consider a feedback control structure
with two controllers, $K^{nl}$ and $K^{lin}$, working in parallel:$K^{nl}$
is a nonlinear controller used to guide the system (\ref{ss_sys})
along the trajectories of interest, while $K^{lin}$ is a linear controller
aimed to enhance the tracking precision.

\subsection{Nonlinear controller design}

The first step needed to design the nonlinear controller is to identify
from the data (\ref{eq:data}) a model for the system (\ref{ss_sys})
of the form 
\begin{equation}
\begin{array}[t]{l}
\hat{y}_{t+1}=f\left(\boldsymbol{y}_{t},\boldsymbol{u}_{t}\right)\equiv f\left(\boldsymbol{q}_{t},u_{t}\right)\vspace{1.5mm}\\
\boldsymbol{q}_{t}=\left(y_{t},\ldots,y_{t-n+1},u_{t-1},\ldots,u_{t-n+1}\right)
\end{array}\label{eq:model}
\end{equation}
where $u_{t}$ and $y_{t}$ are the system input and output, and $\hat{y}_{t}$
is the model output. A parametric structure is taken for the function
$f$: 
\begin{equation}
f\left(\boldsymbol{q}_{t},u_{t}\right)=\sum_{i=1}^{N}\alpha_{i}\phi_{i}\left(\boldsymbol{q}_{t},u_{t}\right)\label{eq:bfe}
\end{equation}
where $\phi_{i}:\mathbb{R}^{n(n_{y}+n_{u})}\rightarrow\mathbb{R}$
are polynomial basis functions and $\alpha_{i}\in\mathbb{R}^{n_{y}\times1}$
are parameter vectors that can be identified by means of convex optimization.

Once a model of the form (\ref{eq:model}) has been identified, the
current command action $u_{t}^{nl}$ of the nonlinear controller $K^{nl}$
is computed by the on-line inversion of the obtained models, given
the available regressor $\boldsymbol{q}_{t}$. This inversion is performed
by solving the following optimization problem: 
\begin{equation}
u_{t}^{nl}=\arg\min_{\bar{u}\in\mathbf{U}}{J_{t}(\bar{u})}\label{ottimizzazione}
\end{equation}
where the current objective function $J_{t}$ is 
\begin{equation}
J_{t}(\bar{u})=\sum_{i=1}^{n_{y}}{\frac{\zeta_{i}}{\rho_{y_{i}}}\Big(r_{t+1,i}-f_{i}(\boldsymbol{q}_{t},\bar{u})\Big)^{2}}+{\frac{\mu_{i}}{\rho_{u_{i}}}\bar{u}^{2}}+{\frac{\lambda_{i}}{\rho_{u_{i}}}\delta\bar{u}^{2}}\label{J}
\end{equation}
where $r_{t+1,i}$ is the $i$th component of $r_{t+1}$ and $f_{i}$
is the $i$th component of $f$; $\rho_{y_{i}}\doteq\frac{||\tilde{y_{i}}_{1-L},\dots,\tilde{y_{i}}_{0}||_{2}^{2}}{L}$
and $\rho_{u_{i}}\doteq\frac{||\tilde{u_{i}}_{1-L},\dots,\tilde{u_{i}}_{0}||_{2}^{2}}{L}$
are normalization constants computed form the data set (\ref{eq:data}),
$\delta\bar{u}=\bar{u}-u_{t-1}$ are the control action rate of change
with respect to the past input, $\mu_{i}\geq0$ and $\lambda_{i}\geq0$
are design parameters which allow us to set the trade off between
tracking precision and control action aggressiveness for every input
(control action magnitude and speed, respectively), and $1\geq\zeta_{i}\geq0$
are other design parameters which are needed to set the priority of
tracking for all the different output. The controller $K^{nl}$ is
fully defined by the law (\ref{ottimizzazione}).

\subsection{Linear controller design}

The linear controller $K^{lin}$ is defined by the centralized extended
PID (Proportional Integrative Derivative) control law 
\begin{equation}
u_{t}^{lin}=u_{t-1}^{lin}+\sum_{i=0}^{n_{\theta}}{B_{i}e_{t-i}}\label{pid}
\end{equation}
where $e_{t}=r_{t}-y_{t}$ is the tracking error, $n_{\theta}$ is
the controller order and $B_{i}\in\mathbb{R}^{{n_{u}\times n_{y}}}$
are the controller parameter matrices. All the entries of these matrices
are contained in a vector $\theta=(\theta_{1},\dots,\theta_{{n_{y}\times n_{u}\times n_{\theta}}}).$
Note that, for $n_{\theta}=1$ and $n_{\theta}=2$, the standard PI
and PID controller are selected, respectively. The goal of $K^{lin}$
in the proposed architecture is to compensate for model-inversion
errors and boost the control performance by assigning a desired dynamics
to the resulting nonlinearly-compensated system.

In the considered setting, where most of the information about the
system is that inferrable from data, finding a good control-oriented
model of the error system, i.e. the system describing the relationship
between $u_{t}^{lin}$ and $y_{t}$, is not an easy task. Therefore,
in this paper, the Virtual Reference Feedback Tuning (VRFT) method,
originally developed in \cite{campi2002virtual} and extended to the
MIMO case in \cite{formentin2012non}, is employed and adapted to
the present setting. Its rationale is briefly recalled next for self-consistency
of the paper.

First, define the desired behavior for the closed loop system by means
of a discrete transfer function $M$. In the present setting, this
function is used to assign a specific desired dynamic to the nonlinearly-compensated
resulting system. Typically, for $n_{y}=n_{u}=m$, $M$ is a diagonal
$m\times m$ transfer function composed by $m$ asymptotically stable
low-pass filters, synthesizing the desired closed loop behavior for
each output. The virtual reference rationale permits the design of
$K^{lin}$ without identifying any model of the system, based on the
following observation: in a virtual operating condition, where the
closed-loop system behaves exactly as $M$, the virtual reference
signal $r_{t}^{v}$ would be given by the filtering of the output
$y_{t}$ by the model $M^{-1}$. Since the inverse model $M^{-1}$
is non-causal, the filtering task must be accomplished off-line using
a set of available data. The optimal controller is the one giving
the measured $u_{t}^{lin}$ as output when fed by the virtual error
${e}_{t}^{v}={r}_{t}^{v}-y_{t}$, as to minimize the cost function
\begin{equation}
J_{VR}=\sum_{t}||u_{t}^{lin}-K^{lin}(\theta){e}_{t}^{v}||_{2}^{2}.\label{minimo5mimovrft}
\end{equation}
Two versions of the VRFT method in the D$^{2}$-IBC setting are available,
corresponding to two cases, related to bijectivity of the function
$f$ in (\ref{eq:model}). The details of these two versions are not
discussed here for the sake of brevity.

\section{Closed-loop stability analysis}

\label{sec:analysis}

The closed-loop system formed by the plant (\ref{ss_sys}), controlled
in feedback by the parallel connection of $K^{nl}$ and $K^{lin}$,
is described by the following equations: 
\begin{equation}
\begin{array}{l}
y_{t+1}=g\left(\boldsymbol{y}_{t},\boldsymbol{u}_{t},\boldsymbol{\xi}_{t}\right)\vspace{1.5mm}\\
u_{t}=u_{t}^{nl}+u_{t}^{lin}\vspace{1.5mm}\\
u_{t}^{nl}=K^{nl}\left(r_{t+1},\boldsymbol{y}_{t},\boldsymbol{u}_{t-1}^{nl}\right)\vspace{1.5mm}\\
u_{t}^{lin}=K^{lin}\left(\boldsymbol{r}_{t}-\boldsymbol{y}_{t},\boldsymbol{u}_{t-1}^{lin}\right)
\end{array}\label{eq:fb}
\end{equation}
where $K^{nl}$ and $K^{lin}$ are defined in (\ref{ottimizzazione})
and (\ref{pid}), respectively, and $u_{t}\in U$, $\forall t$. The
reference initial condition is chosen as $\boldsymbol{r}_{0}=\boldsymbol{y}_{0}$.

In the following, we study the stability properties of such a close-loop
system. The following stability notion is adopted. \smallskip{}
\begin{definition}A nonlinear system (possibly time-varying), with
inputs $r_{t}$ and $\xi_{t}$, and output $y_{t}$, is finite-gain
$\ell_{\infty}$ stable on $\left(\mathcal{Y}^{0},\mathcal{R},\varXi\right)$
if finite and non-negative constants $\Gamma_{r}$, $\Gamma_{\xi}$
and $\Lambda$ exist such that 
\[
\left\Vert \boldsymbol{y}\right\Vert _{\infty}\leq\Gamma_{r}\left\Vert \boldsymbol{r}\right\Vert _{\infty}+\Gamma_{\xi}\left\Vert \boldsymbol{\xi}\right\Vert _{\infty}+\Lambda
\]
for any $(\boldsymbol{y}_{0},\boldsymbol{r},\boldsymbol{\xi})\in\mathcal{Y}^{0}\times\mathcal{R}\times\varXi$.$\qquad\square$
\end{definition}\smallskip{}

Note that this finite-gain stability definition is more general than
the standard one, which is obtained for $\mathcal{R}=\ell_{\infty}$
and $\varXi=\ell_{\infty}$, see e.g. \cite{Khalil96}.

In order to study how to guarantee finite-gain stability of the feedback
system (\ref{eq:fb}), some additional assumptions are introduced.
\smallskip{}
\begin{assumption}[Lipschitzianity]\label{ass:lip_b}The function
$g$ in (\ref{ss_sys}) and (\ref{eq:fb}) is Lipschitz continuous
on $Y^{n}\times U^{n}\times\Xi^{n}$. Without loss of generality,
it is also assumed that $Y^{n}\times U^{n}\times\Xi^{n}$ contains
the origin.$\qquad\square$ \end{assumption} \smallskip{}

\textcolor{black}{This assumption is mild, since most real-world dynamic
systems are described by functions that are Lipschitz continuous on
a compact set.} 

From Assumption \ref{ass:lip_b}, it follows that $g$ can be written
as 
\[
g\left(\boldsymbol{y}_{t},\boldsymbol{u}_{t},\boldsymbol{\xi}_{t}\right)=g^{o}\left(\boldsymbol{y}_{t},\boldsymbol{u}_{t}\right)+g_{t}^{\xi}\boldsymbol{\xi}_{t}
\]
where $g^{o}\left(\boldsymbol{y}_{t},\boldsymbol{u}_{t}\right)\doteq g\left(\boldsymbol{y}_{t},\boldsymbol{u}_{t},\mathbf{0}\right)$
and $g_{t}^{\xi}\in\mathbb{R}^{n_{y}\times n}$ is a time-varying
parameter (dependent on $\boldsymbol{y}_{t}$, $\boldsymbol{u}_{t}$
and $\boldsymbol{\xi}_{t}$) bounded on $Y^{n}\times U^{n}\times\Xi^{n}$
as $\left\Vert g_{t}^{\xi}\right\Vert _{\infty}\leq\gamma_{\xi}$,
for some $\gamma_{\xi}<\infty$. Assumption \ref{ass:lip_b}, together
with (\ref{eq:bfe}), implies that the residue function 
\[
\varDelta\left(\boldsymbol{y}_{t},\boldsymbol{u}_{t}\right)\doteq g^{o}\left(\boldsymbol{y}_{t},\boldsymbol{u}_{t}\right)-f\left(\boldsymbol{y}_{t},\boldsymbol{u}_{t}\right)
\]
is Lipschitz continuous on $Y^{n}\times U^{n}$. Hence, a finite and
non-negative constant $\gamma_{y}$ exists, such that 
\[
\left\Vert \varDelta\left(y,u\right)-\varDelta\left(y',u\right)\right\Vert _{\infty}\leq\gamma_{y}\left\Vert y-y'\right\Vert _{\infty}
\]
for all $y,y'\in Y^{n}$ and all $u\in U^{n}$. \smallskip{}
\begin{assumption}[Model accuracy]\label{ass:lip_delta}The following
inequality holds: $\gamma_{y}<1$.$\qquad\square$ \end{assumption}
\smallskip{}

The meaning of this assumption is clear: it requires $f$ to accurately
describe the variability of $g$ with respect to $\boldsymbol{y}_{t}$. 

Now, consider that
\[
\begin{array}{l}
\hat{e}_{t+1}\doteq r_{t+1}-\hat{y}_{t+1}=r_{t+1}-f\left(\boldsymbol{y}_{t},\boldsymbol{u}_{t}\right)\vspace{1.5mm}\\
\equiv r_{t+1}-f\left(\boldsymbol{y}_{t},\boldsymbol{u}_{t}\left(r_{t+1},\boldsymbol{r}_{t},\boldsymbol{y}_{t},\boldsymbol{u}_{t-1}^{nl},\boldsymbol{u}_{t-1}^{lin}\right)\right)\vspace{1.5mm}\\
\doteq F\left(\boldsymbol{y}_{t},\boldsymbol{r}_{t},\boldsymbol{v}_{t}\right),
\end{array}
\]
where $\boldsymbol{v}_{t}\doteq(r_{t+1},\boldsymbol{u}_{t-1}^{nl},\boldsymbol{u}_{t-1}^{lin})\in V$
and $V$ is a compact set. Then, for any $\left(\boldsymbol{y}_{t},\boldsymbol{r}_{t},\boldsymbol{v}_{t}\right)\in Y^{n}\times R^{n}\times V^{n}$,
\begin{equation}
\left|\hat{e}_{t+1}\right|\leq\Gamma_{y}\left\Vert \boldsymbol{y}_{t}\right\Vert _{\infty}+\Gamma_{s}\left\Vert \boldsymbol{r}_{t}\right\Vert _{\infty}+\Lambda_{e}\label{eq:bbe}
\end{equation}
where $\Gamma_{y},\Gamma_{s},\Lambda_{e}<\infty$. This inequality
directly follows from the fact that the model function $f$ is Lipschitz
continuous on $Y^{n}\times U^{n}$. Note that (\ref{eq:bbe}) does
not imply that $\boldsymbol{y}\in Y^{\infty}$. \smallskip{}
\begin{assumption}[Effective model inversion]\label{ass:clf_stab}
The following inequality holds: $\Gamma_{y}\leq1-\gamma_{y}$.$\qquad\square$\end{assumption}
\smallskip{}

This assumption is not restrictive: it is certainly satisfied if $\mu=0$
and the reference $\boldsymbol{r}=(r{}_{1},r_{2},\ldots)$ is a model
solution (i.e. $r_{t+1}$ is in the range of $f\left(\boldsymbol{y}_{t},\cdot\right)$
for all $t$). Indeed, in this case, $\hat{y}_{t+1}=r_{t+1}$, $\forall t$,
since $K^{nl}$ performs an exact inversion of the model, see again
(\ref{ottimizzazione}) ($K^{lin}$ gives a null input signal in this
case). This implies that $\Gamma_{y}=0$, $\Gamma_{s}=0$ and $\Lambda_{e}=0$.
Hence, if a sufficiently small $\mu$ is chosen and the reference
is sufficiently close to a system solution, supposing that inequality
(\ref{eq:bbe}) holds with a sufficiently small $\Gamma_{y}$ is reasonable. 

To formulate our last assumption, define 
\begin{equation}
\bar{e}\doteq\frac{1}{1-\lambda_{y}}\left(\lambda_{r}\bar{r}+\gamma_{\xi}\bar{\xi}+\Lambda_{g}\right)\label{eq:e_bar}
\end{equation}
where $\lambda_{y}\doteq\Gamma_{y}+\gamma_{y}<1$, $\lambda_{r}\doteq\lambda_{y}+\Gamma_{s}$
and $\Lambda_{g}\doteq\Lambda_{e}+\max_{u\in U^{n}}\left\Vert \varDelta\left(\mathbf{0},u\right)\right\Vert _{\infty}$.
Note that $\bar{e}$ is bounded, being the sum of bounded quantities.
In particular, for null disturbances ($\bar{\xi}=0$), exact modeling
($f=g$, $\Delta=0$, $\gamma_{y}=0$) and reference signals properly
chosen ($\Gamma_{y}=0$, $\Gamma_{s}=0$, $\Lambda_{e}=0$), we have
$\bar{e}=0$. In realistic situations, with reasonable disturbances,
sufficiently accurate modeling and reference signals properly chosen,
$\bar{e}$ can be reasonably small (that is, $\bar{e}\ll\bar{r}$).\smallskip{}
\begin{assumption}[Output domain exploration]\label{ass:clf_stab-1}
The following inequality holds: $\bar{y}\geq\bar{r}+\bar{e}$.$\qquad\square$\end{assumption}
\smallskip{}

This assumption requires that the set $Y$ explored by the output
data is somewhat larger than the set $R$ where the trajectory of
interest are defined. Note that it can always be met just collecting
data that sufficiently enlarge the set $Y$.

Closed-loop stability of the system (\ref{eq:fb}) is stated by the
following result, which also provides a bound on the tracking error.
\smallskip{}
\begin{theorem}\label{thm:stab}Consider the system (\ref{eq:fb})
and let Assumptions \ref{ass:lip_b}-\ref{ass:clf_stab-1} hold. Then:\\
(i) The feedback system (\ref{eq:fb}), having inputs $r_{t}$ and
$\xi_{t}$ and output $y_{t}$, is finite-gain $\ell_{\infty}$ stable
on $\left(\mathcal{Y}^{0},\mathcal{R},\varXi\right)$. \\
(ii) The tracking error signal $\boldsymbol{e}\doteq\boldsymbol{r}-\boldsymbol{y}$
is bounded as 
\begin{equation}
\left\Vert \boldsymbol{e}\right\Vert _{\infty}\leq\bar{e}.\label{eq:teb}
\end{equation}
\end{theorem}

\textbf{Proof.} The proof of the theorem is structured as follows.
Firstly, the tracking error is proven to be upper bounded by a suitable
combination of the norms of the output and the reference. Secondly,
it is shown that, under the assumption of an effective model inversion,
such a bound is equivalent to a bound on the tracking error, whatever
the output is (claim (ii)). Claim (i) is derived as a straightforward
consequence of claim (ii).

To start with, consider that 
\[
e_{t+1}\doteq r_{t+1}-y_{t+1}=\hat{e}_{t+1}-\delta y_{t}
\]
where 
\[
\begin{array}[t]{l}
\hat{e}_{t+1}=r_{t+1}-\hat{y}_{t+1}=F\left(\boldsymbol{y}_{t},\boldsymbol{r}_{t},\boldsymbol{v}_{t}\right)\vspace{1.5mm}\\
\delta y_{t}=\varDelta\left(\boldsymbol{y}_{t},\boldsymbol{u}_{t}\right)+g_{t}^{\xi}\boldsymbol{\xi}_{t}.
\end{array}
\]
The term $\hat{e}_{t+1}$ is bounded according to (\ref{eq:bbe}).
Note that $\left\Vert \boldsymbol{y}_{t}\right\Vert _{\infty}$ could
be unbounded. In order to derive a bound on $\delta y_{t}$, we can
use Assumption \ref{ass:lip_delta} and observe that, for any $\boldsymbol{y}_{t}\in Y^{n}$,
\[
\begin{array}[t]{c}
\left\Vert \varDelta\left(\boldsymbol{y}_{t},\boldsymbol{u}_{t}\right)\right\Vert _{\infty}-\left\Vert \varDelta\left(\mathbf{0},\boldsymbol{u}_{t}\right)\right\Vert _{\infty}\vspace{1.5mm}\\
\leq\left\Vert \varDelta\left(\boldsymbol{y}_{t},\boldsymbol{u}_{t}\right)-\varDelta\left(\mathbf{0},\boldsymbol{u}_{t}\right)\right\Vert _{\infty}\leq\gamma_{y}\left\Vert \boldsymbol{y}_{t}\right\Vert _{\infty}.
\end{array}
\]
The following inequality thus holds for any $\boldsymbol{y}_{t}\in Y^{n}$:
\[
\begin{array}{c}
\left\Vert \delta y_{t}\right\Vert _{\infty}\leq\left\Vert \varDelta\left(\boldsymbol{y}_{t},\boldsymbol{u}_{t}\right)\right\Vert _{\infty}+\gamma_{\xi}\left\Vert \boldsymbol{\xi}_{t}\right\Vert _{\infty}\vspace{1.5mm}\\
\leq\gamma_{y}\left\Vert \boldsymbol{y}_{t}\right\Vert _{\infty}+\gamma_{\xi}\left\Vert \boldsymbol{\xi}_{t}\right\Vert _{\infty}+\bar{\Delta},
\end{array}
\]
where $\bar{\Delta}\doteq\max_{u\in U^{n}}\left\Vert \varDelta\left(\mathbf{0},u\right)\right\Vert _{\infty}<\infty.$
Hence, 
\begin{equation}
\begin{array}{c}
\left\Vert e_{t+1}\right\Vert _{\infty}\leq\lambda_{y}\left\Vert \boldsymbol{y}_{t}\right\Vert _{\infty}+\Gamma_{s}\left\Vert \boldsymbol{r}_{t}\right\Vert _{\infty}+\Lambda_{e}\vspace{1.5mm}\\
+\gamma_{\xi}\left\Vert \boldsymbol{\xi}_{t}\right\Vert _{\infty}+\bar{\Delta},
\end{array}\label{eq:etb0}
\end{equation}
which proves that the tracking error $e$ is bounded by a suitable
combination of the norms of $\boldsymbol{y}_{t}$ and $\boldsymbol{r}_{t}$,
for any $\boldsymbol{y}_{t}\in Y^{n}$. Note that (\ref{eq:etb0})
in this form is of no use, since $\left\Vert \boldsymbol{y}_{t}\right\Vert _{\infty}$
could be unbounded and thus the condition $\boldsymbol{y}_{t}\in Y^{n}$
may not hold. However, (\ref{eq:etb0}) can be rewritten as 
\begin{equation}
\begin{array}{c}
\left\Vert e_{t+1}\right\Vert _{\infty}\leq\lambda_{y}\left\Vert \boldsymbol{e}_{t}\right\Vert _{\infty}+\lambda_{y}\left\Vert \boldsymbol{r}_{t}\right\Vert _{\infty}+\Gamma_{s}\left\Vert \boldsymbol{r}_{t}\right\Vert _{\infty}\vspace{1.5mm}\\
+\gamma_{\xi}\left\Vert \boldsymbol{\xi}_{t}\right\Vert _{\infty}+\Lambda_{g},
\end{array}\label{eq:etb01}
\end{equation}
that means, 
\begin{equation}
\begin{array}{c}
\left\Vert e_{t+1}\right\Vert _{\infty}\leq\lambda_{y}\left\Vert \boldsymbol{e}_{t}\right\Vert _{\infty}+w,\end{array}\label{eq:etb02}
\end{equation}
where $w\doteq\lambda_{r}\bar{r}+\gamma_{\xi}\bar{\xi}+\Lambda_{g}$.
Inequality (\ref{eq:etb02}), again, holds only if $\boldsymbol{y}_{t}\in Y^{n}$.\\
 Consider now that, by assumption, $\boldsymbol{y}_{0}\in R^{n}\subseteq Y^{n}$.
This implies that, at time $t=0$, inequality (\ref{eq:etb02}) holds.
Being $\boldsymbol{e}_{0}=\boldsymbol{r}_{0}-\boldsymbol{y}_{0}=0$
for the selected initialization of $\boldsymbol{r}_{0}$, we have
\[
\left\Vert e_{1}\right\Vert _{\infty}\leq\lambda_{y}\left\Vert \boldsymbol{e}_{0}\right\Vert _{\infty}+w=w\leq\bar{e}.
\]
Since $\left\Vert e_{1}\right\Vert _{\infty}\leq\bar{e}$ and $\left|r_{1}\right|\leq\bar{r}$,
it follows from Assumption \ref{ass:clf_stab-1} that $y_{1}\in Y$.
Consequently, the Lipschitzianity assumption holds and (\ref{eq:etb02})
can be used also for $t=1$, giving 
\[
\begin{array}{c}
\left\Vert e_{2}\right\Vert _{\infty}\leq\lambda_{y}\left\Vert e_{1}\right\Vert _{\infty}+w\leq\lambda_{y}w+w\vspace{1.5mm}\\
\leq w\sum_{k=0}^{\infty}\lambda_{y}^{k}\leq\frac{w}{1-\lambda_{y}}=\bar{e}
\end{array}
\]
where the geometric series sum has been obtained thanks to the fact
that, by Assumption \ref{ass:clf_stab}, $\lambda_{y}<1$. It follows
that $y_{2}\in Y$ and, consequently, (\ref{eq:etb02}) can be used
also for $t=2$. Iterating the above reasoning, 
\[
\begin{array}{c}
\left\Vert e_{3}\right\Vert _{\infty}\leq\lambda_{y}\max\left\{ \left\Vert e_{2}\right\Vert _{\infty},\left\Vert e_{1}\right\Vert _{\infty}\right\} +w\vspace{1.5mm}\\
\leq\lambda_{y}\left\Vert e_{2}\right\Vert _{\infty}+w\leq\lambda_{y}^{2}w+\lambda_{y}w+w\leq\bar{e}\\
\vdots\\
\left\Vert e_{t+1}\right\Vert _{\infty}\leq w\sum_{k=0}^{t}\lambda_{y}^{k}\vspace{1.5mm}\\
\leq w\sum_{k=0}^{\infty}\lambda_{y}^{k}\leq\frac{w}{1-\lambda_{y}}=\bar{e}.
\end{array}
\]
Then, $y_{t}\in Y$, $\forall t\geq0$ and (\ref{eq:teb}) holds (claim
(ii)). Claim (i) is a direct consequence of claim (ii) and the relation
$\boldsymbol{y}=\boldsymbol{r}-\boldsymbol{e}$.$\qquad\square$ 

\bibliographystyle{IEEEtran}
\bibliography{lettnos_journals,lettnos_conferences,sparsification,intelligent_bib,lettnos_others,simo_biblio}

\end{document}